\documentclass[prl,twocolumn,floatfix,preprintnumbers,letterpaper]{revtex4}
\usepackage{graphicx}
\usepackage{bm}
\usepackage{epsfig}
\usepackage{amsmath,amsfonts}
\usepackage{url}

\begin{document}

\title{Thawing $f(R)$ cosmology}
\author{Mahmood Roshan and Fatimah Shojai}
\affiliation{Department of Physics, University of Tehran, Tehran,
Iran }

\begin{abstract}
We consider Brans-Dicke (BD) scalar tensor theory in the conformally
transformed Einstein frame. In this frame BD theory behaves like an
interacting quintessence model. We find the necessary conditions on
the form of the potential $V(\varphi)$ in order to have thawing
behavior. Finally, by setting the BD coupling constant $\omega=0$,
the metric $f(R)$ gravity has been considered in the Einstein frame.
Assuming the existence of thawing solution, some necessary
conditions for $f(R)$ gravity models have been derived.
\end{abstract}
\maketitle
\section{Introduction}
One of the proposals for explaining the present accelerated
expansion of the universe \cite{sahni} is modifying Einstein's
theory of gravity by introducing corrections to the
Einstein-Hilbert lagrangian. These theories, called "modified
gravity theories" \cite{ds} follow this idea that the accelerated
expansion of the universe may be has a geometric interpretation
instead of adding the exotic forms of energy sources, dubbed "dark
energy" \cite{sami}. In the other words, in this perspective the
dark energy is a manifestation of a modified gravitational
interaction rather than a new form of energy density. The
situation is reminiscent of the problem of precession of Mercury's
orbit. In the mid-nineteenth century the anomalous behavior of
Mercury firstly was attributed to some unobserved ("dark") planet
in the solar system while it was mainly due to the failure of
Newton's theory of gravity in the strong gravitational field
regime. In this view, it seems that as long as the dark energy
particles \cite{dark particles} have not been observed directly,
the "geometric" candidates have important role. The simplest form
of the modified gravity theories can be obtained by replacing the
Ricci scalar $R$ with an arbitrary general function $f(R)$ in the
Einstein-Hilbert action, usually called $f(R)$ theory of gravity.
For a recent review of this theory see \cite{faraoni}.

Metric $f(R)$ gravity model is dynamically equivalent to a BD
scalar tensor theory with coupling constant $\omega=0$
\cite{sotiriou}. By using this equivalence, one can easily find
the prediction of metric $f(R)$ gravity for the PPN parameter
$\gamma_{PPN}$. This parameter in the BD theory has the form
$\gamma_{PPN}=(1+\omega)/(2+\omega)$. Thus the value of this
parameter in the metric $f(R)$ gravity is $1/2$ which it is not in
agreement with the experimental bound
$|\gamma_{PPN}-1|<2.3:10^{-5}$\cite{bertotti}. However,
considering this model in the Einstein frame has some satisfactory
features. For example, $f(R)$ gravity can display the chameleon
behavior in this frame which helps to relax the weak field limit
problem of $f(R)$ gravity \cite{chameleon f}. Chameleon effect is
firstly interpreted using the scalar tensor framework of dark
energy \cite{khoury}. In this theory the effective mass of the
scalar field is a function of the curvature of space-time and
consequently it can be large at the solar system and small on the
cosmological scales. This behavior appears in the minimally
coupled scalar tensor theory if there exists an energy transfer
between the dark energy fluid and the ordinary matter fluid. Since
the quintessence model \cite{quintessence} is a minimally coupled
scalar tensor theory, the chameleon mechanism can be appeared. On
the other hand, metric $f(R)$ gravity theories are conformally
equivalent to models of quintessence in which matter is coupled to
the dark energy, thus the chameleon effect can occur in the
conformal frame \cite{chameleon f}.

 The noninteracting quintessence models can be divided into two categories
\cite{linder}. "Freezing" models: in these models the equation of
state parameter of dark energy, $\omega_{\varphi}$, has an arbitrary
value initially and decreases with time and asymptotically
approaches $-1$. "Thawing" models: these models have a value of
$\omega_{\varphi}\sim-1$ initially, and it increases with time.
There is a subset of freezing models which display tracking behavior
\cite{steinhardt}. In the tracking models, $\omega_{\varphi}$ has an
arbitrary value initially and it is nearly constant during the
tracking era. When the tracking era terminates then
$\omega_{\varphi}$ decreases and asymptotically approaches -1. The
important feature of these models is that the evolution of the
scalar field is insensitive to the initial conditions and the dark
energy density drops with a slower rate than the matter energy
density and finally overtakes it. Albeit, these models can not
provide a solution to the so-called coincidence problem because
other fine-tunings are needed on the free parameters of these models
in order to have an appropriate amount of dark energy compatible
with observation in the present days\cite{coincidence}.

In our recent paper\cite{we} we derived some conditions for
existing the stable tracker solutions in the Einstein frame of
metric $f(R)$ gravity models. It is found that the tracker
solutions with $-0.361<\omega_{\varphi}<1$ exist if
$0<\Gamma<0.217$ and $\frac{d}{dt}\ln f'(\tilde{R})>0$, where
$\Gamma$ is a dimensionless function defined by relation
\eqref{Gamma} in the next section. The main purpose of this paper
is to find out the necessary conditions for the existence of
thawing behavior in the Einstein frame of metric $f(R)$ gravity
theories.

The outline of this paper is as follows: In section II we start
with BD scalar tensor theory (with an arbitrary $\omega$). As
mentioned before, this theory behaves as an interacting
quintessence model in this frame. We derive some necessary
conditions on the form of the potential $V(\varphi)$ in order to
lead to the thawing behavior for $\omega_{\varphi}$. In section
III, by setting $\omega=0$ in the results, we present a general
description of the behavior of the thawing $f(R)$ in the Einstein
frame and finally conditions on the form of $f(R)$ gravity have
been derived. Throughout this work we have chosen the unit $8\pi
G=c=1$, the metric signature is $(+---)$ and the universe is
assumed to be spatially flat.

\section {Thawing nonminimal quintessence}
The effective action for BD scalar tensor theory is given by
\begin{equation}
S_{J}=\int\sqrt{-\tilde{g}}\ d^{4}x \
[\Phi\tilde{R}-\frac{\omega}{\Phi}\Phi^{,\mu}\Phi_{,\mu}-2U(\Phi)+\mathcal{L}_{m}(\tilde{g}_{\mu\nu})]
\end{equation}
where $\tilde{R}$ is the Ricci scalar, $U(\Phi)$ is the potential
of the scalar field and $\mathcal{L}_{m}(\tilde{g}_{\mu\nu})$
represents the matter lagrangian density. Note that all tilded
quantities are in the Jordan frame. The coupling constant $\omega$
should be large to pass the experimental testes. The observational
constraint on $\omega$ is $|\omega|>40000$ \cite{bertotti}. Under
the conformal transformation
\begin{equation}\label{conft}
g_{\mu\nu}=e^{\zeta\varphi}\tilde{g}_{\mu\nu}
\end{equation}
where $\ln \Phi=\zeta\varphi$ and
$\zeta=\sqrt{\frac{2}{3+2\omega}}$, one can obtain the Einstein
frame action
\begin{equation}\label{einstein action}
\begin{split}
S_{E}=\int\sqrt{-g} \ d^{4}x\
[R-\frac{1}{2}\varphi^{,\mu}&\varphi_{\mu}-V(\varphi)\\&+\mathcal{L}_{m}(g_{\mu\nu}e^{-\zeta\varphi})]
\end{split}
\end{equation}
where $V(\varphi)=e^{-2\zeta \varphi}U(\Phi(\varphi))$. We see
that in the Einstein frame the scalar field couples conformally to
matter via the function $e^{-\zeta\varphi}$ but couples minimally
to the gravity sector. For a spatially flat FRW universe, the
modified Friedmann equations are given by
\begin{equation}\label{feridmann}
\begin{split}
&H^{2}=\frac{1}{3}(\rho_{\varphi}+\rho_{m})\\
&\dot{H}=-\frac{1}{2}[(1+\omega_{m})\rho_{m}+(1+\omega_{\varphi})\rho_{\varphi}]
\end{split}
\end{equation}
and the equation of motion of the scalar field is
\begin{equation}\label{scalar field eom}
\ddot{\varphi}+3H\dot{\varphi}+V_{\varphi}=\sqrt{\frac{2}{3}}\beta(1-3\omega_{m})\rho_{m}
\end{equation}
where $\beta=\sqrt{\frac{3}{8}}\ \zeta$, $\omega_{m}$ is the
equation of state parameter of the ordinary matter with the energy
density $\rho_{m}$ in the Einstein frame. Also
$\rho_{\varphi}=\frac{1}{2}\dot{\varphi}^{2}+V(\varphi)$ and
$p_{\varphi}=\frac{1}{2}\dot{\varphi}^{2}-V(\varphi)$ represent
the energy density and pressure of the dark energy respectively.
The conservation equations of the scalar field fluid and the
cosmic fluid are
\begin{equation}\label{conservation}
\begin{split}
&\dot{\rho}_{\varphi}+3H(1+\omega_{\varphi})\rho_{\varphi}=\sqrt{\frac{2}{3}}\beta\dot{\varphi}(1-3\omega_{m})\rho_{m}\\
&\dot{\rho}_{m}+3H(1+\omega_{m})\rho_{m}=-\sqrt{\frac{2}{3}}\beta\dot{\varphi}(1-3\omega_{m})\rho_{m}
\end{split}
\end{equation}
and the energy density of matter $\rho_{m}$, pressure $p_{m}$,
cosmic time $t$ and the scale factor $a$ are related to their
Jordan frame counterparts through \cite{faraonibook}
\begin{eqnarray}
\rho_{m}=e^{-2\zeta\varphi}\tilde{\rho}_{m},
p_{m}=e^{-2\zeta\varphi}\tilde{p}_{m},
dt=e^{\frac{\zeta\varphi}{2}}d\tilde{t},
a=e^{\frac{\zeta\varphi}{2}}\tilde{a}
\end{eqnarray}
During the matter dominated era, by using equation (\ref{scalar
field eom}), one can introduce an effective potential as follows
\begin{eqnarray}\label{Veffective}
V_{eff}(\varphi)=V(\varphi)+\rho^{*}e^{-\sqrt{\frac{2}{3}}\beta\varphi}
\end{eqnarray}
Where $\rho^{*}$ is a conserved quantity in the Einstein frame
\cite{khoury}, which is related to $\rho_{m}$ via the relation
$\rho_{m}=\rho^{*}e^{-\sqrt{\frac{2}{3}}\beta\varphi}$.

Since the late time evolution of the universe is of interest here
and also our main purpose is to explore the role of the
interaction term (which is nonzero for the matter component), we
neglect the radiation component and assume that the universe
contains only dust and dark energy. It is interesting to note that
the interaction term is commonly assumed to be zero in the
radiation dominated era, but recently Cembranos and \textit{et al}
\cite{cem} have shown that this interaction term can lead to
strong impact on cosmology in the radiation dominated era due to
the finite temperature radiative corrections. In the other words,
there exists another source term for scalar field given by the
conformal anomaly which leads to a nonzero trace of energy
momentum tensor in the radiation dominated era (note that the RHS
of \eqref{scalar field eom} is the trace of energy momentum
tensor). Considering the conformally coupled scalar field with a
quadratic coupling function and vanishing potential, the above
effect leads to a temporary contracting phase in which the
temperature increases\cite{cem}. However, as mentioned before, we
aim to study here the late time evolution of the universe and so
we assume that the universe is filled with non-relativistic
matter.

Following reference \cite{copeland}, we introduce the variables $x$,
$y$ and $\lambda$ defined by
\begin{eqnarray}
x=\frac{\varphi'}{\sqrt{6}}, \ \ \
y=\sqrt{\frac{V(\varphi)}{3H^{2}}}, \ \ \
\lambda=-\frac{V_{\varphi}}{V}
\end{eqnarray}
where the prime denotes the derivative with respect to $\ln a$. By
these definitions, it is an easy job to show that the equations
\eqref{feridmann} and \eqref{scalar field eom} become
\begin{eqnarray}\label{x,y}
\begin{split}
&x'=-3x+\lambda\sqrt{\frac{3}{2}}y^{2}+\frac{3}{2}x(1+x^{2}-y^{2})+\beta(1-x^{2}-y^{2})\\
&y'=-\lambda\sqrt{\frac{3}{2}}x y+\frac{3}{2}y(1+x^{2}-y^{2})\\
&\lambda'=-\sqrt{6}\lambda^{2}(\Gamma-1)x
\end{split}
\end{eqnarray}
where
\begin{eqnarray}\label{Gamma}
\Gamma=V\frac{d^{2}V}{d\varphi^{2}}/(\frac{d V}{d\varphi})^{2}
\end{eqnarray}
For thawing models $\omega_{\varphi}\sim -1$ and so
$\gamma=1+\omega_{\varphi}\ll1$. Thus it is convenient to express
the above equations with respect to $\gamma$ in order to exploit
its smallness by expanding quantities to the lowest order in
$\gamma$. Also we assume that $\dot{\varphi}>0$ $(x'>0)$. This
assumption, as considered in \cite{we}, is necessary to have an
increasing dark energy density parameter i.e.
$\dot{\Omega}_{\varphi}>0$. However, the results can be
generalized to the opposite case $(x'<0)$. Now, by using
$\Omega_{\varphi}=x^{2}+y^{2}$ and
$\gamma=2x^{2}/\Omega_{\varphi}$, one can rewrite the equations
\eqref{x,y} in terms of $\gamma$ and $\Omega_{\varphi}$ as
\begin{eqnarray}\label{gammaprime}
\gamma'=(2-\gamma)\left(-3\gamma+\lambda\sqrt{3\gamma\Omega_{\varphi}}+\sqrt{\frac{2\gamma}{\Omega_{\varphi}}}\beta(1-\Omega_{\varphi})\right)
\end{eqnarray}
\begin{eqnarray}\label{omegaprime}
\Omega_{\varphi}'=3(1-\Omega_{\varphi})\left((1-\gamma)\Omega_{\varphi}+\frac{\beta}{3}\sqrt{2\gamma\Omega_{\varphi}}\right)
\end{eqnarray}
\begin{eqnarray}
\lambda'=-\sqrt{3}\lambda^{2}(\Gamma-1)\sqrt{\gamma\Omega_{\varphi}}
\end{eqnarray}
It is clear from equation \eqref{omegaprime} that for thawing
models $\Omega'_{\varphi}\neq0$ during the cosmological history of
the universe $(0<\Omega_{\varphi}<1)$. Thus by using
\eqref{omegaprime} we can write equation \eqref{gammaprime} as
follows
\begin{eqnarray}\label{go}
\frac{d\gamma}{d\Omega_{\varphi}}=\frac{(2-\gamma)\left(-3\gamma+
\lambda\sqrt{3\gamma\Omega_{\varphi}}+\sqrt{\frac{2\gamma}{\Omega_{\varphi}}}\beta(1-\Omega_{\varphi})\right)}
{3\Omega_{\varphi}(1-\Omega_{\varphi})\left(1-\gamma+\frac{\beta}{3}\sqrt{\frac{2\gamma}{\Omega_{\varphi}}}\right)}
\end{eqnarray}
This equation is obtained earlier in \cite{sen} in which, the
non-minimal quintessence with nearly flat potentials has been
considered. Equation \eqref{go} is not a simple differential
equation and for solving it, we will make some assumptions which
are satisfied for thawing models. First assume that $\gamma\ll1$,
by retaining terms up to the first order in $\gamma$, the equation
\eqref{go} takes the following form
\begin{eqnarray}\label{go1}
\begin{split}
\frac{d\gamma}{d\Omega_{\varphi}}&\simeq
\frac{-2\gamma}{\Omega_{\varphi}(1-\Omega_{\varphi})}-\sqrt{\frac{8}{21}}\beta\frac{\lambda\gamma}{\Omega_{\varphi}(1-\Omega_{\varphi})}
\\&+\frac{2\lambda\sqrt{\gamma}}{\sqrt{3\Omega_{\varphi}}(1-\Omega_{\varphi})}
+
\frac{2\beta\sqrt{2\gamma}}{3\Omega_{\varphi}^{3/2}}-\frac{4\beta^{2}\gamma}{9\Omega_{\varphi}^{2}}
\end{split}
\end{eqnarray}
Another useful equation can be obtained by using the equation of
motion of the scalar field \eqref{scalar field eom}
\begin{eqnarray}\label{lambda}
\lambda=\sqrt{\frac{3\gamma}{\Omega_{\varphi}}}[1+\frac{\gamma'}{3\gamma(2-\gamma)}]-\frac{\sqrt{\frac{8}{3}}\beta}{2-\gamma}
\frac{1-\Omega_{\varphi}}{\Omega_{\varphi}}
\end{eqnarray}
This equation can be written to the first order in $\gamma$ as
follows
\begin{equation}\label{lambdaa}
\lambda\simeq\left[\sqrt{\frac{3\gamma}{\Omega_{\varphi}}}(1+\frac{\gamma'}{6\gamma})-\sqrt{\frac{1}{6}}\beta\gamma
\frac{1-\Omega_{\varphi}}{\Omega_{\varphi}}\right]-\sqrt{\frac{2}{3}}\beta\frac{1-\Omega_{\varphi}}{\Omega_{\varphi}}
\end{equation}

\begin{figure}[!t]
\hspace{0pt}\rotatebox{0}{\resizebox{0.4\textwidth}{!}{\includegraphics{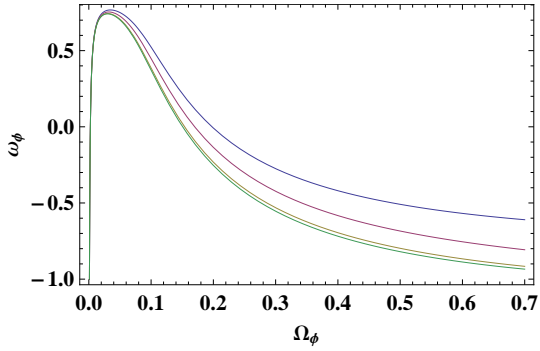}}}
\vspace{0pt}{\caption{Numerical solutions of \eqref{go} for nearly
flat potentials when $\beta=0.5$ for (top to bottom) $\lambda=1$,
$\lambda=0.5$, $\lambda=0.1$ and $\lambda=0.01$. Assume that
$\omega_{\varphi}\simeq-1$ at $\Omega_{\varphi}=0.001$.
}\label{nearly flat}}
\end{figure}

 For uncoupled quintessence, where $\beta$ is zero, the RHS of
equation \eqref{lambda} is approximately constant and moreover it
has an small amount (note that $\gamma\ll1$) for thawing solutions
. Thus if $(\frac{V_{\varphi}}{V})^{2}\ll1$ then the thawing
behavior can occur \cite{scherrer}. In the general case where
$\beta$ is not zero, then the RHS can not be regarded as a
constant. Since $\Omega_{\varphi}$ is appeared in the denominator,
hence the second term in the RHS is dominated initially and has a
large value. Thus the LHS can not have a small value as well as
can not be a constant. When $\Omega_{\varphi}$ gets larger, the
effect of the interaction becomes weaker. Thus, at late times, the
nearly flat region of the potential leads to the thawing uncoupled
quintessence. So, unlike the noninteracting quintessence model,
nearly flat potentials can not lead to the thawing behavior when
an explicit energy transfer between the scalar field fluid and the
matter fluid exists. In this case, as mentioned in \cite{sen},
with nearly flat potentials, $\omega_{\varphi}$ firstly increases
with time and then, when the interaction becomes weaker, it
decreases and approaches asymptotically to a value near $-1$. The
behavior of $\omega_{\varphi}$ with nearly flat potentials has
been plotted in Fig.\ref{nearly flat} by solving equation
\eqref{go} numerically. Note that this behavior is due to the
special form of interaction which appeared here (i.e.
$\dot{\varphi}\rho_{m}$).

Now we are ready to make the second assumption. Taking into account
equation \eqref{lambdaa} and assuming that the value of the term
within the bracket to be approximately constant for thawing
solutions, this equation gives
\begin{equation}\label{lambda1}
\lambda\simeq\lambda_{0}-\sqrt{\frac{2}{3}} \beta
\frac{\rho_{m}}{\rho_{\varphi}}=\lambda_{0}-\sqrt{\frac{2}{3}}
\beta\frac{1-\Omega_{\varphi}}{\Omega_{\varphi}}
\end{equation}
where $\lambda_{0}$ is a positive constant. Hereafter we shall
refer to this equation as the "thawing condition". For potentials
in which $-\frac{V_{\varphi}}{V}$ decreases as $\varphi$ increases
($\Gamma>1$), the LHS of the equation \eqref{lambda1} is
increasing. On the other hand, the RHS is increasing because
$\Omega_{\varphi}$ increases. Hence, the thawing condition can not
be satisfied. Thus, the thawing condition shows that it is
necessary $\lambda$ increases with time when $\varphi$ and
$\Omega_{\varphi}$ are increasing, i.e.
\begin{equation}\label{G}
\Gamma<1
\end{equation}
It is clear from the thawing condition that if $\beta=0$ then
$\lambda$ is nearly constant and so $\Gamma\simeq1$. One can find
other simple conditions on the form of $V(\varphi)$ by using the
thawing condition. For this purpose, let us rewrite \eqref{lambda1}
as follows
\begin{equation}
\frac{1}{V}\frac{dV}{d\varphi}\simeq-(\lambda_{0}+\sqrt{\frac{2}{3}}
\beta)+\sqrt{\frac{2}{3}} \beta\frac{1}{\Omega_{\varphi}}
\end{equation}
It is clear from this equation that the second term in the RHS is
dominated initially and so $\frac{dV}{d\varphi}>0$. As mentioned
before, the interaction becomes weaker at late times. Thus, there
exists a time $t^{*}$, at which
\begin{equation}
\left(\lambda_{0}+\sqrt{\frac{2}{3}}
\beta\right)_{t=t^{*}}\simeq\sqrt{\frac{2}{3}}
\beta\left(\frac{1}{\Omega_{\varphi}}\right)_{t=t^{*}}
\end{equation}
At this time $\frac{dV}{d\varphi}\simeq0$ and after it
$\frac{dV}{d\varphi}<0$. Thus, it is necessary that the potential
has a maximum in order to have thawing behavior. In the other
words, a value of the scalar field, $\varphi^{*}$ should exist
such that
\begin{equation}\label{V}
\begin{split}
&\frac{dV}{d\varphi}|_{\varphi^{*}}=0\\
&\frac{d^{2}V}{d\varphi^{2}}|_{\varphi^{*}}<0
\end{split}
\end{equation}

\begin{figure}[!t]
\hspace{0pt}\rotatebox{0}{\resizebox{0.4\textwidth}{!}{\includegraphics{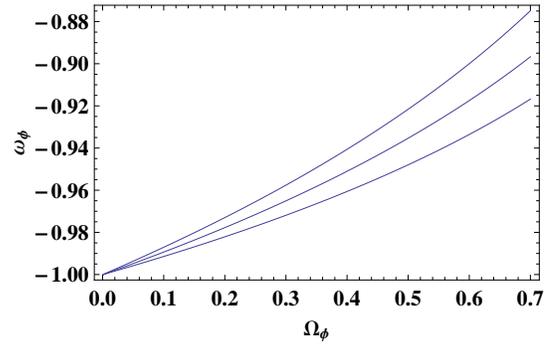}}}
\vspace{0pt}{\caption{Our analytical result for $\omega_{\varphi}$
as a function of $\Omega_{\varphi}$ for $\beta=0.5$ and
$\Omega_{\varphi0}=0.7$ for (top to bottom)$\lambda_{0}=1$,
$\lambda_{0}=0.9$ and $\lambda_{0}=0.8$. Also, as an initial value,
it has been assumed that at $\Omega_{\varphi}=0.001$ $\gamma$ is
zero. ($\Omega_{\varphi0}$ is the current value of
$\Omega_{\varphi}$)}\label{beta0.5}}
\end{figure}

\begin{figure}[!t]
\hspace{0pt}\rotatebox{0}{\resizebox{0.4\textwidth}{!}{\includegraphics{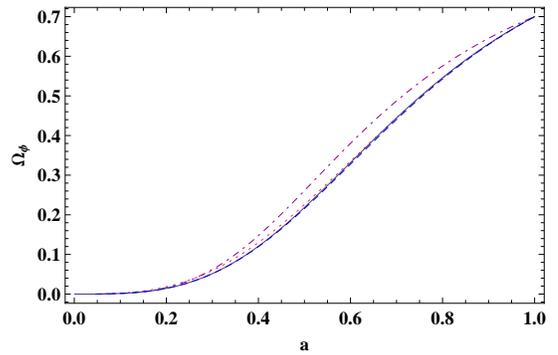}}}
\vspace{0pt}{\caption{The solid curve represent the exact solution
of \eqref{omegaprime} when $\gamma=0$, i.e. the equation \eqref{om},
assuming $\Omega_{\varphi0}=0.7$. The dot dashed curve is the
numerical solution of \eqref{omegaprime} with $\lambda_{0}=2$, the
dotted curve is for $\lambda_{0}=1$ and the dashed curve is for
$\lambda_{0}=0.8$.}\label{numericallomega}}
\end{figure}

 Now let us to justify the thawing condition. By these assumptions ($\gamma\ll1$ and \eqref{lambda1}), equation \eqref{go1} takes the very
simple following form
\begin{eqnarray}\label{go2}
\frac{d\gamma}{d\Omega_{\varphi}}+\frac{2A\gamma}{\Omega_{\varphi}(1-\Omega_{\varphi})}-
\frac{2\lambda_{0}\sqrt{\gamma}}{\sqrt{3\Omega_{\varphi}}(1-\Omega_{\varphi})}\simeq0
\end{eqnarray}
in which $A=1+\beta\sqrt{\frac{2}{27}}\lambda_{0}$. Note that if
$\beta=0$, then this equation becomes precisely the equation
obtained earlier by Scherrer and Sen \cite{scherrer}. This
differential equation has an exact solution as follows
\begin{eqnarray}\label{finalgamma}
\begin{split}
\gamma=\left(\frac{1-\Omega_{\varphi}}{\Omega_{\varphi}}\right)^{2A}[\
\chi_{0}+&\frac{2\lambda_{0}\Omega_{\varphi}
^{1/2+A}}{\sqrt{3}(1+2A)}\ \\
&_{2}F_{1}(\frac{1}{2}+A,1+A,\frac{3}{2}+A,\Omega_{\varphi})\ ]^{2}
\end{split}
\end{eqnarray}
where $\chi_{0}$ is an integration constant depending on the
initial conditions and $_{2}F_{1}$ is the Gauss Hypergeometric
function. This equation gives an analytical expression for the
state parameter of dark energy as a function of its density
parameter for the thawing non-minimal quintessence model. It
generalizes the result obtained in \cite{scherrer} for thawing
minimally coupled scalar field. The behavior of $\omega_{\varphi}$
as a function of $\Omega_{\varphi}$ has been shown in
Fig.\ref{beta0.5} for various values of $\lambda_{0}$. The $\beta$
has been chosen to be $0.5$, the value will be used in the next
section in the case of $f(R)$ gravity models. In fact, the value
of $\lambda_{0}$ should be such that $\omega_{\varphi}$ has a
value near $-1$ today.
 For confronting the model with observational data,
it is needed to express $\gamma$ and $\Omega_{\varphi}$ in terms
of cosmic red shift or cosmic scale factor. By substituting the
solution \eqref{finalgamma} in the equation \eqref{omegaprime}, we
obtain a differential equation for $\Omega_{\varphi}$ in which the
Gauss Hypergeometric function is appeared. Here, we have solved it
numerically and the result has been compared with the exact
solution of the equation \eqref{omegaprime} when $\gamma=0$, in
Fig.\ref{numericallomega}. Thus as it is clear from
Fig.\ref{numericallomega}, the difference between these solutions
is small when $\lambda_{0}\sim1$ and consequently one can use the
solution of \eqref{omegaprime} when $\gamma=0$ i.e.
\begin{equation}\label{om}
\Omega_{\varphi}=[1+(\Omega_{\varphi0}^{-1}-1)a^{-3})]^{-1}
\end{equation}
in order to find out an approximated expression for $\gamma$ as a
function of $a$.

\section{Thawing $f(\tilde{R})$}
Now let us consider $f(\tilde{R})$ gravity in the Einstein frame.
It is sufficient to set $\omega=0$ (and so
$\zeta=\sqrt{\frac{2}{3}}$) in equation \eqref{finalgamma} in
order to have thawing behavior in the Einstein frame. Also in the
context of $f(\tilde{R})$ gravity, the scalar field $\varphi$ is
related to the curvature scalar of the Jordan frame as follows
\begin{equation}\label{conf}
\begin{split}
&\varphi=\sqrt{3/2}\ln f_{\tilde{R}}(\tilde{R})\\
&V(\varphi)=(\tilde{R}f_{\tilde{R}}-f)/2f_{\tilde{R}}^{2}
\end{split}
\end{equation}
where $f_{\tilde{R}}=\frac{df}{d\tilde{R}}$. As it is clear from
Fig.\ref{beta0.5}, $\lambda_{0}$ has been chosen near to $1$ in
order to have $\omega_{\varphi}$ near $-0.9$. In this case, as it
has been shown in Fig.\ref{numericallomega}, $\Omega_{\varphi}$
evolves as the dark energy density parameter of $\Lambda$CDM model
in which $\omega_{\varphi}$ is always approximated to $-1$.

Now, we want to find out explicit conditions on the form of the
function $f(\tilde{R})$ in order to have thawing behavior in the
Einstein frame. Taking into account equations \eqref{G} and
\eqref{conf}, one can easily verify that if
$f_{\tilde{R}\tilde{R}}>0$ then
\begin{equation}\label{f1}
\tilde{R}f_{\tilde{R}}^{2}-(\tilde{R}f_{\tilde{R}\tilde{R}}+f_{\tilde{R}})f<0
\end{equation}
and for $f_{\tilde{R}\tilde{R}}<0$
\begin{equation}\label{f2}
\tilde{R}f_{\tilde{R}}^{2}-(\tilde{R}f_{\tilde{R}\tilde{R}}+f_{\tilde{R}})f>0
\end{equation}
Also, by using \eqref{V} and \eqref{conf}, it is necessary that
the form of $f(\tilde{R})$ be such that there exists
$\tilde{R}^{*}$ for which
\begin{equation}\label{f3}
\begin{split}
&\tilde{R}^{*}=\frac{2f}{f_{\tilde{R}}}|_{\tilde{R}^{*}}\\
&\tilde{R}^{*}>\frac{f_{\tilde{R}}}{f_{\tilde{R}\tilde{R}}}|_{\tilde{R}^{*}}
\end{split}
\end{equation}
Note that for having nonsingular conformal transformation (equations
\eqref{conf} and \eqref{conft}) we have assumed $f_{\tilde{R}}>0$.
Equations \eqref{f1}-\eqref{f3} are the necessary conditions on the
form of $f(\tilde{R})$ for raising to the thawing behavior and they
are not sufficient conditions.

\begin{figure}[!t]
\hspace{0pt}\rotatebox{0}{\resizebox{0.4\textwidth}{!}{\includegraphics{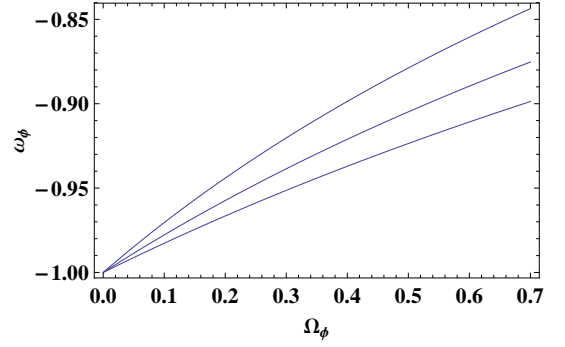}}}
\vspace{0pt}{\caption{$\omega_{\varphi}$ as a function of
$\Omega_{\varphi}$ for $\beta=0.5$ for (top to bottom)$\alpha=2/7$,
$\alpha=1/4$ and $\alpha=2/9$.}\label{example}}
\end{figure}

Now, let us to find out an explicit example for thawing potentials.
For this purpose, assume that
\begin{equation}\label{1}
\frac{d\varphi}{d\Omega_{\varphi}}=\frac{\alpha}{1-\Omega_{\varphi}}
\end{equation}
where, $\alpha$ is a positive constant. This assumption leads to
the following form of dark energy density parameter
\begin{equation}\label{2}
\Omega_{\varphi}=1-e^{-\frac{(\varphi-\psi)}{\alpha}}
\end{equation}
which is an increasing function of $\varphi$ and $\psi$ is an
integration constant. Using
$\varphi'=\sqrt{3\gamma\Omega_{\varphi}}$ and equations
\eqref{omegaprime}, \eqref{1} and \eqref{2}, we obtain
\begin{equation}\label{3}
\gamma=\frac{B+18\alpha^{2}\Omega_{\varphi}\mp\sqrt{B^{2}+36B\alpha^{2}\Omega_{\varphi}}}{18\alpha^{2}\Omega_{\varphi}}
\end{equation}
where $B=3-2\sqrt{2}\alpha\beta+2\alpha^{2}\beta^{2}$. If $\alpha$
is a small quantity ($\alpha<1$), then the solution with minus
sign can yield to the thawing behavior. For seeing this,
$\omega_{\varphi}$ has been plotted in Fig.\ref{example} for
various values of $\alpha$. By substituting equation \eqref{3}
into equation \eqref{lambda} and expanding the RHS of equation
\eqref{lambda} to the second order in $\alpha$ (note that we have
assumed that $\alpha$ is small), we get
\begin{equation}\label{4}
\frac{1}{V}\frac{dV}{d\Omega_{\varphi}}\approx\frac{-\sqrt{\frac{2}{3}}\beta}{\Omega_{\varphi}}\alpha+\frac{1+6\Omega_{\varphi}}
{2(1-\Omega_{\varphi})}\alpha^{2}+O(\alpha^{3})
\end{equation}
which has the following solution
\begin{equation}\label{5}
V\approx
V_{0}\Omega_{\varphi}^{\sqrt{\frac{2}{3}}\alpha\beta}(1-\Omega_{\varphi})^{\frac{7\alpha^{2}}{2}}e^{3\alpha^{2}\Omega_{\varphi}}
\end{equation}
\begin{figure}[!t]
\hspace{0pt}\rotatebox{0}{\resizebox{0.4\textwidth}{!}{\includegraphics{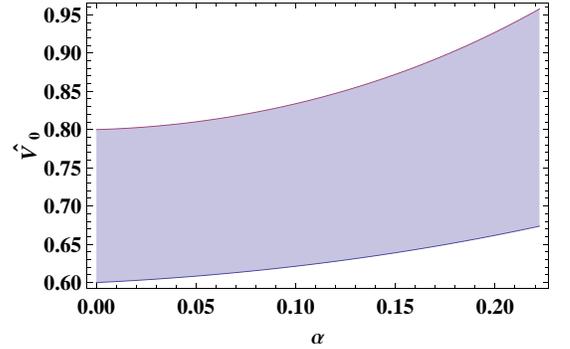}}}
\vspace{0pt}{\caption{Region of parameter space compatible with the
observational constraints $-1<\omega_{\varphi 0}<-0.9$ and
$0.6\leq\Omega_{\varphi0}\leq0.8$ for
$\beta=0.5$.}\label{fintuning}}
\end{figure}
where $V_{0}$ is a positive integration constant. It is obvious from
this that $V$ has a maximum and so it is consistent with our
pervious results. By setting $\psi$ to be zero and using equation
\eqref{2}, let us rewrite equation \eqref{5} as a function of the
scalar field as follows
\begin{equation}\label{6}
V\approx V_{0} \
e^{-\frac{7\alpha}{2}\varphi}(1-e^{-\frac{\varphi}{\alpha}})^{\sqrt{\frac{2}{3}}\alpha\beta}
e^{3\alpha^{2}(1-e^{-\frac{\varphi}{\alpha}})}
\end{equation}
This potential satisfies the conditions \eqref{G} and \eqref{V} and
it is a two parameter potential ($V_{0}$ and $\alpha$). The
parameter $\alpha$ should be small and for $-1<\omega_{\varphi
0}<-0.9$ it should be $0<\alpha<0.23$. Thus, the only free parameter
in this model is $V_{0}$. As mentioned before, this free parameter
should be fin-tuned by using the observational data. The
observational fact is that the energy density of dark energy and the
energy density of cosmic matter fluid are approximately in the same
order. Since the potential has been obtained with respect to
$\Omega_{\varphi}$ (equation \eqref{5}) it is easy to make an
estimation on the values of $V_{0}$ to reproduce the acceleration
expansion. Albeit, we assume that the major contribution to the
energy of the scalar field is due to the potential term (note that
this is the case for all thawing potentials). The density parameter
of dark energy is
\begin{equation}\label{7}
\Omega_{\varphi}\sim\frac{V}{3H^{2}}
\end{equation}
By using this equation and \eqref{5} we obtain
\begin{equation}\label{7}
\frac{V_{0}}{3H_{0}^{2}}\sim\Omega_{\varphi0}^{1-\sqrt{\frac{2}{3}}\alpha\beta}
(1-\Omega_{\varphi0})^{-\frac{7\alpha^{2}}{2}}e^{-3\alpha^{2}\Omega_{\varphi0}}
\end{equation}
where $H_{0}$ is the current value of the Hubble parameter. By
taking into account that the current value of $\Omega_{\varphi0}$
satisfies the bound $0.6\leq\Omega_{\varphi0}\leq0.8$, we have
plotted the region of parameter space able to cover the above
observational constraints, in Fig. \ref{fintuning}. This region
varies from $\alpha=0$ to $\alpha=0.23$ and from
$\hat{V_{0}}\approx0.6$ to $\hat{V_{0}}\approx0.95$, where
$\hat{V_{0}}$ is a dimensionless variable defined as follows
\begin{equation}
\hat{V_{0}}=\frac{V_{0}}{3H_{0}^{2}}
\end{equation}
Note that for making a more precise estimation one should use
numerical solutions of the field equations and taking into account
the effect of the kinetic term of energy density of the dark energy,
see the third paper of \cite{coincidence} and also \cite{fin} for
more details.

Now, for finding the corresponding $f(\tilde{R})$ function, assume
that $f(\tilde{R})$ differs from Einstein's general relativity by
a small perturbation as follows
\begin{equation}\label{f}
f(\tilde{R})=\tilde{R}+\varepsilon \Psi(\tilde{R})
\end{equation}
where $\varepsilon$ is a very small parameter. By substituting
this in equation \eqref{conf}, using \eqref{6} and taking
$\beta=0.5$, one reaches to the following first order differential
equation up to the first order in $\varepsilon$
\begin{equation}\label{df2}
-\varepsilon
\Psi(\tilde{R})+\varepsilon\Psi_{\tilde{R}}\approx2V_{0}\left(\frac{3}{2}\right)^{\frac{\alpha}{2\sqrt{6}}}
\left(\frac{\varepsilon\Psi_{\tilde{R}}}{\alpha}\right)^{\sqrt{\frac{1}{6}}\alpha}
\end{equation}
which has the solution of the form
\begin{equation}\label{df3}
\varepsilon\Psi(\tilde{R})\approx-\mu\tilde{R}^{n}
\end{equation}
where
\begin{equation}
\begin{split}
&n=\frac{\alpha}{\alpha-\sqrt{6}}\\
&\mu=2V_{0}(1-\frac{\alpha}{\sqrt{6}})\left(\frac{12}{\sqrt{9^{\alpha}}V_{0}}\right)^{n}
\ 3^{\frac{\alpha}{\sqrt{6}}}
\end{split}
\end{equation}
Since $\alpha$ is a positive constant, $n$ is a negative real
number. Thus, the perturbation procedure is valid if the curvature
of space time is sufficiently high such that $\left|-\mu
\tilde{R}^{n}\right|\ll\tilde{R}$. Consequently, the model
\eqref{df3} can lead to the thawing behavior only in the beginning
of matter dominated era. However, for larger values of $\varphi$,
let us expand the potential \eqref{6} again. Before proceeding, we
expect that $\Psi(\tilde{R})$ contains some terms of $\tilde{R}$
with powers smaller than $n$. Because such a term can have effect in
the late times (large $\varphi$), where the curvature is small,
while it can be neglected compared with $\tilde{R}^{n}$ where the
curvature is larger.

If $\varphi$ is large enough such that
$e^{-\frac{\varphi}{\alpha}}\ll1$ then one can write the potential
\eqref{6} as follows
\begin{equation}\label{df5}
V\approx V_{0}e^{3\alpha^{2}} \ e^{-\frac{7\alpha}{2}\varphi}
\end{equation}
It is easy to show that the $f(\tilde{R})$ function corresponding
to this potential is
\begin{equation}\label{df6}
f(\tilde{R})\sim\nu\tilde{R}^{m}
\end{equation}
where
\begin{equation}
\begin{split}
&m=\frac{7\sqrt{6}\alpha-8}{7\sqrt{6}\alpha-4}\\&
\nu=\frac{2V_{0}e^{3\alpha^2}(1-7/2\sqrt{3/2}\alpha)}{(2V_{0}e^{3\alpha^2}(2-7/2\sqrt{3/2}\alpha))^{m}}
\end{split}
\end{equation}
It is possible to make an estimation on the value of $\alpha$ by
assuming that the potential \eqref{df5} is a solution of
differential equation \eqref{lambda1} when
$\Omega_{\varphi}\sim1$. Thus, $\lambda_{0}\sim\frac{7}{2}\alpha$
and consequently $\alpha\sim\frac{2}{7}$. By this amount for
$\alpha$, $m$ is negative and also it is smaller than $n$ (m=\
-3.45), as we expected.

As a result, the following model
\begin{equation}\label{final}
f(\tilde{R})\sim\tilde{R}-\mu\tilde{R}^{n}+\nu\tilde{R}^{m}
\end{equation}
can lead to the thawing behavior in the matter dominated epoch and
late times. Note that this model satisfies the condition \eqref{G}.
Also, as we required, it's corresponding potential in the Einstein
frame has a maximum.

\section{DISCUSSION}
We have considered BD scalar tensor theory in the Einstein frame.
In this frame, BD theory behaves like an interacting quintessence.
It is necessary that the potential $V(\varphi)$ has a maximum in
the region where the scalar field rolls in order to have thawing
behavior. Also the potential should satisfy the condition
$\Gamma<1$. The thawing condition \eqref{lambda1} shows that for
non-interacting quintessence model, potential should satisfy the
condition $\Gamma\approx1$ \cite{scherrer}.

In the last section, by setting the BD coupling constant $\omega$
to zero, we have studied the thawing behavior of $f(\tilde{R})$
gravity models in the Einstein frame. It is important to note that
for power law $f(\tilde{R})$ gravity models, such as \eqref{df6},
the equation of state parameter of dark energy (in the Einstein
frame) firstly increases with time and then decreases. This
behavior is due to the form of the corresponding potential of
these models in the Einstein frame. For these models, $\lambda$ is
constant and as we have mentioned in section II,
$\omega_{\varphi}$ evolve as Fig. \ref{nearly flat}. The sign of
the power of $\tilde{R}$ depends on the present day value of
$\omega_{\varphi}$ and it's magnitude depends on $\lambda_{0}$. As
it is clear from Fig. \ref{nearly flat}, choosing different values
for $\lambda_{0}$ leads to different values of $\omega_{\varphi}$
at the present day.

As an example for thawing $f(\tilde{R})$ models, we have proposed
the model given by \eqref{final}. The corresponding potential has a
maximum and the condition $\Gamma<1$ is satisfied. So, in the
beginning of the matter dominated era $\lambda$ is not approximately
constant, (see equation \eqref{4}), and $\omega_{\varphi}$ increases
slowly as it has been shown in Fig. \ref{example}. Also this model
leads to a nearly flat potential in the late times which is
satisfactory. At sufficiently late times, the interaction term in
equation \eqref{lambda} is negligible and we expect our model
behaves like a non-interacting thawing quintessence \cite{scherrer}.

 \section{acknowledgments}

We would like to thank the referee for useful comments. M. Roshan
would like to thanks J.A. Cembranos for useful hints and
communications. This work is partly supported by a grant from
university of Tehran and partly by a grant from center of
excellence of department of physics on the structure of matter.

\end{document}